\documentclass[a4,showpacs,twocolumn, amsmath, amssymb, pre, aps, floatfix, superscriptaddress]{revtex4-1}
\usepackage{epsfig}
\usepackage{hyperref}
\usepackage{graphicx,color}
\usepackage{amsmath}
\usepackage{dcolumn}
\usepackage{multirow}
\usepackage[normalem]{ulem}
\begin{document}

\title{Lamellar and inverse micellar structures of 
skin lipids: Effect of templating} 
 \author{Chinmay Das}
 \email{c.das@leeds.ac.uk}
 \affiliation{School of Physics and Astronomy, University of Leeds,
 Leeds LS2 9JT, United Kingdom}
 \author{Massimo G. Noro}
 \email{Massimo.Noro@unilever.com}
 \affiliation{Unilever R$\&$D, Port Sunlight, Wirral, CH63 3JW, United Kingdom}
 \author{Peter D. Olmsted}
 \email{p.d.olmsted@leeds.ac.uk}
 \affiliation{School of Physics and Astronomy, University of Leeds,
 Leeds LS2 9JT, United Kingdom}
\date{\today}

\begin{abstract}
The outermost layer of skin comprises
rigid non-viable cells (corneocytes) in a layered lipid matrix.
Using atomistic simulations we find that 
the equilibrium phase of the skin lipids is 
inverse micellar.
A model of the corneocyte is used to demonstrate that lamellar
layering is induced by the patterned corneocyte wall. 
The inverse micellar phase is consistent with {\em in vivo} observations
in regions where corneocyte walls are well separated (lacunar spaces) and 
in the inner layers of skin,
and suggests a functional role in the lipid synthesis pathway {\em in vivo}. 
\end{abstract}
\maketitle
\begin{figure}[htbp]
\centerline{\includegraphics[width=\linewidth]{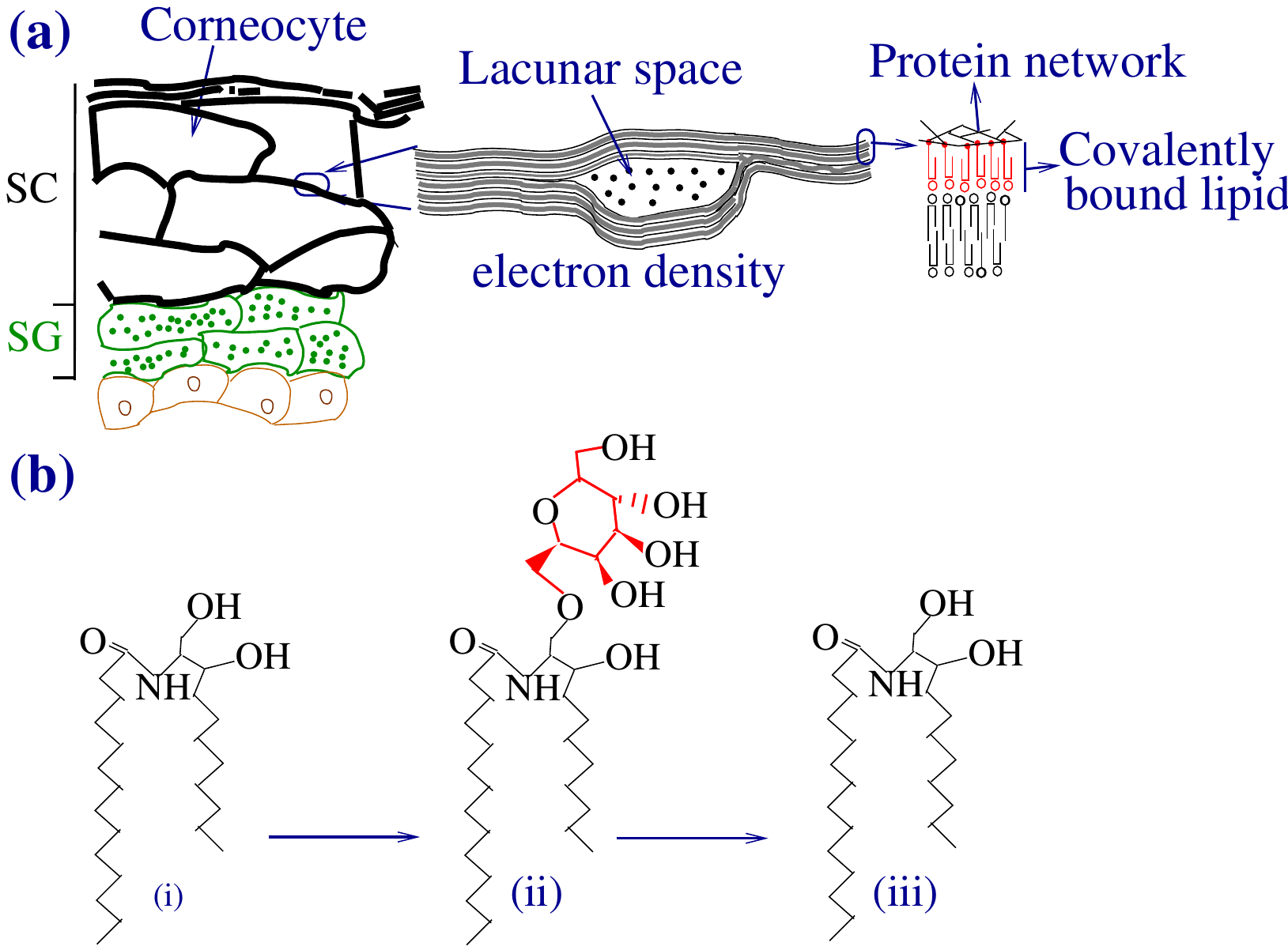}}
\caption{(color online) (a) Schematic representation of the top two layers of skin:
SC, the outermost layer, contains rigid corneocytes in a
lipid matrix. 
Between corneocytes, the lipids
show a periodic electron density pattern that is lost when
the corneocyte walls are further apart \protect{\cite{amoudi.cubic.cryoem.05}}.
A layer of lipids are covalently bound to the
protein network of corneocytes forming the 
`corneocyte bound lipid envelope' (CLE). 
(b) CER is synthesised in SG, the layer immediately below SC. 
After synthesis, a sugar moeity is added
to the head group and the resulting galactoceramides are released 
by endocytosis. Enzymes remove the
sugar group to revert back to CER in the intercellular space
\protect{\cite{Mizutani.09}}.}  
\label{fig.sc.intro}
\end{figure}

{\bf Introduction:}
The outermost layer of skin, the stratum corneum (SC), comprises
non-viable rigid cells (mainly keratin filled corneocytes) in a 
lipid matrix (Fig.~\ref{fig.sc.intro}a). 
The continuous lipid matrix is responsible for the extraordinary
barrier property of skin and is the first line of defence against 
invasion of foreign pathogens \cite{elias.sc.rev.05}. The integrity of the lipid matrix
is essential for proper functioning of skin, 
and an understanding of its structures will help
in the design of agents to rejuvenate damaged lipid layers and
selectively perturb the lipid structure to temporarily and
reversibly increase the permeability for trans-dermal drug delivery
\cite{prausnitz.04}.

The SC matrix is conspicuous 
by the absence of lipids with polar head groups and by having large molecular 
polydispersity. 
The earliest models introduced a coarse picture of ``bricks'' (corneocyte) 
surrounded by 
``mortar'' (lipid) \cite{michaels.sc.brick.75}.
Later {\em in~vivo} observations 
\cite{breathnach.freezefracture.73, white.biochem.88, bouwstra.91, amoudi.cubic.cryoem.05},  and 
experiments with reconstituted SC lipid layers \cite{groen.bba.08,schroter.bpj.09}, showed 
a multilamellar arrangement
of lipids.
Scattering experiments \cite{white.biochem.88, bouwstra.91, schroter.bpj.09} and cryo-EM 
\cite{breathnach.freezefracture.73, amoudi.cubic.cryoem.05}
images
yielded a large number of apparently contradictory results: 
lipids in gel-like \cite{norlen.jid.01} or fluid phases \cite{bommannan.jid.90}, 
one \cite{amoudi.cubic.cryoem.05} or two \cite{bouwstra.91} periodicities (in a multilayer
arrangement), and signatures of different crystalline arrangements \cite{pilgram.jid.99}. 
Existing models \cite{swartzendruber.jid.89,
bouwstra.jlr.98,forslind.adv.94,hill.bba.03,mcintosh.bpj.03,iwai.jid.12}
 have described some of these features by including 
specific lipids 
in particular positions of 
essentially a periodic crystalline arrangement. 
However, skin employs more than 300
different lipid molecules, and the relative concentrations 
of the different components vary widely (more than 100\% between individuals
and across the body sites of the same individual
\cite{norlen.sccomp.99}) without
affecting the normal skin function. 
A periodic fixed molecular arrangement is unlikely to accommodate such large
variations in composition, or explain
the low but finite permeability for small molecules 
\cite{bartek.permeability.72},
the pliability of healthy skin in accommodating
deformation from mechanical and hydration stresses 
\cite{wu.sc.mechanical.06}, or the activity 
of colocalised antimicrobial peptides and proteases \cite{aberg.amp.07}.

Ceramide sphingolipids (CER) constitute 
$\sim 30 \,\textrm{mol\%}$
of SC lipids \cite{weerheim.sccomp.01, farwanah.sccomp.cer.05}. 
CER is synthesised in the stratum granulosum (SG), 
a layer below the SC \cite{Mizutani.09}. 
Immediately after synthesis a sugar moiety
is attached to the head group 
of the CER (fig.~\ref{fig.sc.intro}b)
and the resulting galactoceramides are secreted by endocytosis.
Enzymes in the extra-cellular space
remove the sugar group to revert the molecules back to CER 
\cite{Mizutani.09}.
In response
to removing the SC lipids 
by chemical insult, cells in the SG release lipid
vesicles within minutes \cite{menon.jid.92}, but the permeability barrier recovers
only after a few days \cite{grubauer.jlr.89}. The seemingly unnecessary step of
attaching and removing a head group, and the large separation
of time-scale between lipid release and barrier recovery, remain
unexplained.

Previous molecular simulations of SC lipids have been limited to 
selected lipid components in pre-formed hydrated bilayers without
realistic polydispersity
\cite{holtje.fachol.01,pandit.cer2.06, notman.dmso.07,
das.bpj.09,das.smat.10,hoopes.jpcb.11,engelbrecht.smat.11}. 
Here we report results from large scale molecular dynamics simulations
( $\sim 4 \times 10^6$ united atoms  and $\mu$s time scales)
 of SC lipids with realistic polydispersity.
The main findings are: (i) Randomly
oriented initial conditions lead to an inverse-micellar
arrangement in 30 wt\%  water. 
(ii) This structure is not just  kinetically trapped,
since an initial lamellar phase structure transforms to an inverted phase 
in simulation time-scales. 
(iii) A weak 
multilamellar arrangement develops when the lipid molecules are confined
between two walls that mimic the corneocyte-bound lipid envelope. 
These results are consistent with
existing {\em in-vivo} and {\em in-vitro} 
observations on the SC lipid arrangments, and can 
shed light on
the relevance of the temporary addition of a head group in ceramide
biosynthesis and the reason for separation of lipid-release and
barrier-recovery timescales.

\begin{figure}[htbp]
\centerline{\includegraphics[width=0.8\linewidth]{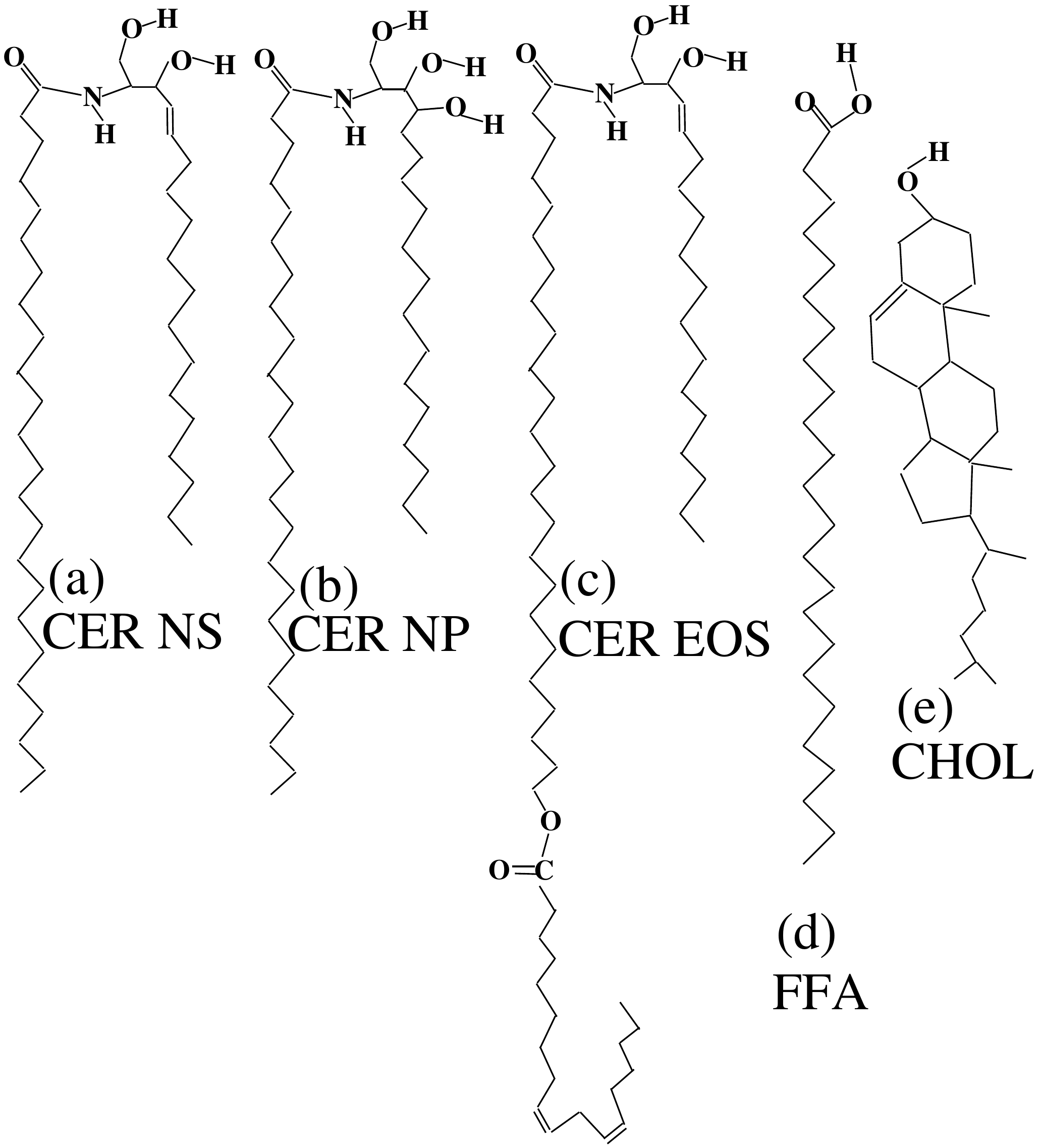}}
\caption{Schematic representation of selected SC lipid molecules:
CER~NP (b) differs from CER~NS (a) by having an extra OH group. 
CER~EOS (c) contains an additional linoleic acid
conjugated to the fatty acid tail. Both CER (a-c) and FFA (d) 
have large polydispersity in the tail lengths.}
\label{fig.lipid.mol}
\end{figure}

{\bf Simulations:}
The three main lipid components of the SC are CER, 
free fatty acids (FFA), and cholesterol (CHOL) 
\cite{weerheim.sccomp.01, farwanah.sccomp.cer.05}. 
By changing the numbers and positions of the hydroxy groups in 
the sphingosine motif, nature 
uses 11 different families of CER molecules - each having 
large polydispersity  in the tail lengths \cite{supmat}.
FFA molecules also show similar polydispersity in the tail lengths.
We use the GROMACS molecular dynamics package \cite{gromacs95}
with the `Berger' force-field \cite{chiu.ff.95,berger.ff.97} for the lipids
and the SPC model for the water molecules. 
To probe the bulk arrangement, for
the 1:1:1 composition (molar ratio of CER, CHOL and FFA), we
use 2000 CER, 2000 CHOL and 2000 FFA molecules and 50000 water
molecules (3.7$\times$10$^6$ united atoms). 
We use three members of
the ceramide family (CER~NS, CER~NP and CER~EOS Fig.\ref{fig.lipid.mol}),
with the experimentally observed tail length polydispersity,
and thus 15 different CER molecules. The molar fractions of
these three families were increased to account for the 
CER molecules of similar structure not explicitly considered in the 
simulations.  
Simulation details are included in 
the online supplementary material \cite{supmat}.

\begin{figure}[htbp]
\includegraphics[width=\linewidth]{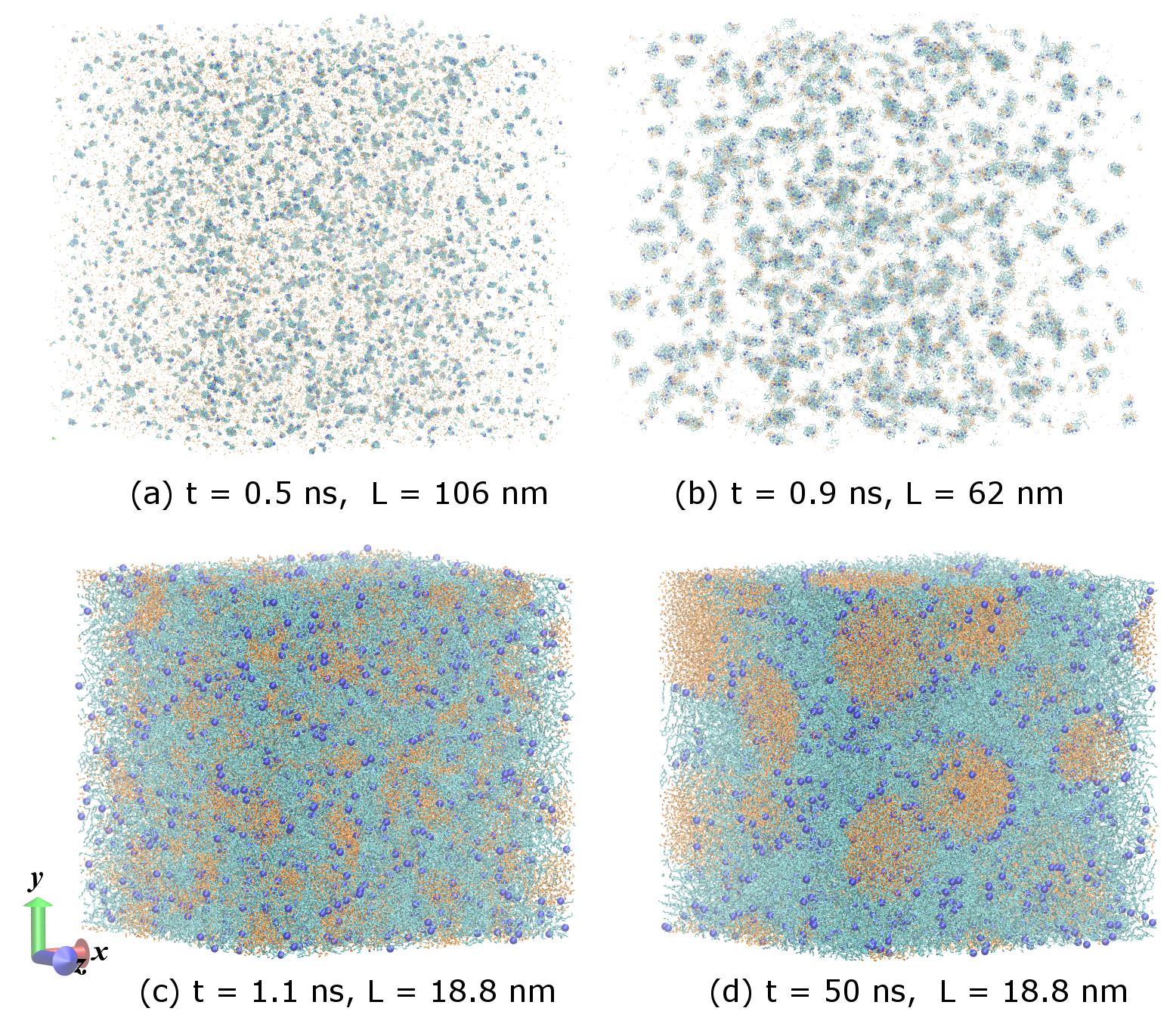}
\caption{Evolution of structure in 1:1:1 molar ratio of CER, CHOL and FFA 
SC lipids (cyan)
with 30 wt.\% water (orange). 
(a): 0.5~ns after random placement of the molecules in a box of
dimension $L=120 \textrm{nm}$, $L$ reduces slightly to $106 \textrm{nm}$. 
Water and lipid molecules start to cluster. (b): At $0.9~\textrm{ns}$ the clustering
is more pronounced and the density increases quickly ($L \sim 62 \textrm{nm}$).
(c): At $1.1 \;\textrm{ns}$ the equilibrium density has been reached
($L \sim 18.8 \textrm{nm}$). The clumping of water is now more pronounced.
(d): By $50 \textrm{ns}$, the water molecules have arranged in a few large
roughly spherical droplets, enclosed by coronae of lipid
molecules \protect{\cite{supmat}}.} 
\label{fig.invmic.evol}
\end{figure}

{\bf Hydrated bulk SC lipids - inverse micelles:}
To probe the bulk structure of the SC lipids in 
30 wt.\% water (approximate water content in the outer SC \cite{warner.sc.hyd.88}, though
much of this water remains inside corneocytes \cite{bouwstra.corneocyte.03})
 with the least 
amount of bias, we place randomly oriented lipid and water molecules 
at random positions in a large simulation box. 
The imposed
atmospheric pressure compresses this `gas' 
until most of the water molecules aggregate 
into a few large and roughly spherical clusters, leading to
an inverse-micellar phase (Fig.~\ref{fig.invmic.evol}). We used three
different initial conditions each for 1:1:1 and 2:2:1 composition ratios
of CER, CHOL and FFA molecules. In all cases the system 
acquired an inverse micellar phase 
\cite{supmat}.  

In one of the 2:2:1 systems we prepared the initial 
simulation box with the $z$-dimension much smaller than the lateral 
dimensions. After the isotropic compression to atmospheric pressure,
but before the water molecules formed isolated clusters, we used 
different imposed pressures along the $z$-direction (1 bar) and the 
lateral directions (1000 bar).
The resulting flow aligns the lipid tails preferentially along the 
$z$-direction. 
Once the box becomes roughly cubic, we switched back to an isotropic pressure
coupling. The final structure, once again, is an inverse micellar arrangement 
with the lipid tails isotropically distributed.  
This shows that liquid-crystalline ordering 
of the lipid molecules alone is not strong enough to sustain a lamellar 
arrangement.

\begin{figure}[htbp]
\includegraphics[width=\linewidth]{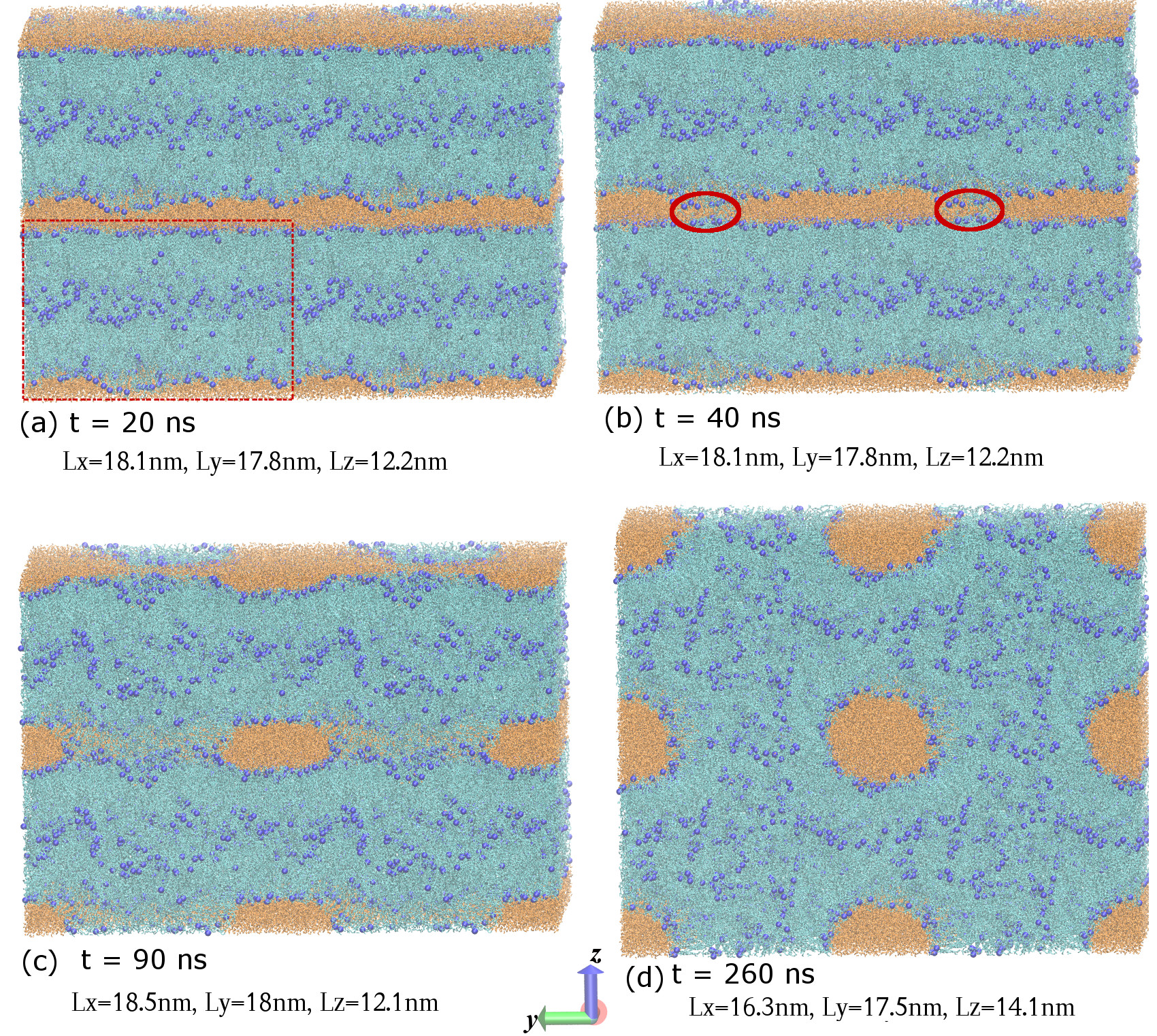}
\caption{SC lipid (cyan) double bilayer (equimolar CER, CHOL and FFA) with
a thin layer of water (orange). There are no water molecules between the two
bilayers. Nitrogen atoms in the head groups are
shown as  blue spheres.
The simulation box is repeated once in both the $y$ and $z$ directions
(The actual box dimension is indicated by dashed rectangle).
(a): Lipid head groups in a small region across the thin water layer
come closer to each other by natural fluctuations at around 20 ns. (b) The
strong interaction between head groups pulls the leaflets closer (circled). 
(c) The water molecules evolve  towards a cylindrical shaped region. 
Substantial rearrangement leads  to a
bilayer corona of lipid molecules around the water cylinder. 
(d) The structure remains
stable for the rest of the simulation (260~ns) \cite{supmat}.} 
\label{fig.mult.unstable}
\end{figure}

To further probe whether or not the inverse micellar phase is due to the 
preparation process,
we created a dry double bilayer (1:1:1 composition).
After equilibration with a continuous water wall along the $z$-direction,
the continuous water wall was replaced by a thin layer of water
($\sim 1 \textrm{nm}$) and standard periodic boundary conditions were
used in all three directions. 
When, by natural
fluctuation, a few of the lipid molecules in the outer leaflet of
the top bilayer come in contact with the outer leaflet of the bottom bilayer
across the thin water layer (via periodic boundary conditions), 
the head-head contact area increases irreversibly, 
 creating
columns of water separated by the lipid corona (Fig.~\ref{fig.mult.unstable}).
The lipids retain a bilayer topology, with a cylindrical water column
surrounded by {\em bilayers} instead of the usual monolayers.  

A small fully hydrated bilayer of SC lipids is
indefinitely stable in simulations. The local stress tensor
across such a bilayer contains information about the spontaneous
curvature favoured by the molecules \cite{seddon.90}. Using
local pressure tensor data from a 2:2:1 bilayer simulation, we find 
the spontaneous curvature $c_0$ for (half of) the bilayer as
$1/c_0 \sim - 15\,\textrm{nm}$ \cite{supmat}. 
The sign of the curvature points towards
equilibrium inverse phases and the magnitude is similar to the
length-scale of the observed checkerboard pattern at 
the SC-SG boundary in cryo-EM \cite{amoudi.cubic.cryoem.05}.

\begin{figure}[htbp]
\hbox{\vbox{\parbox{0.49\linewidth}{
\centerline{\includegraphics[width=0.95\linewidth,clip=]{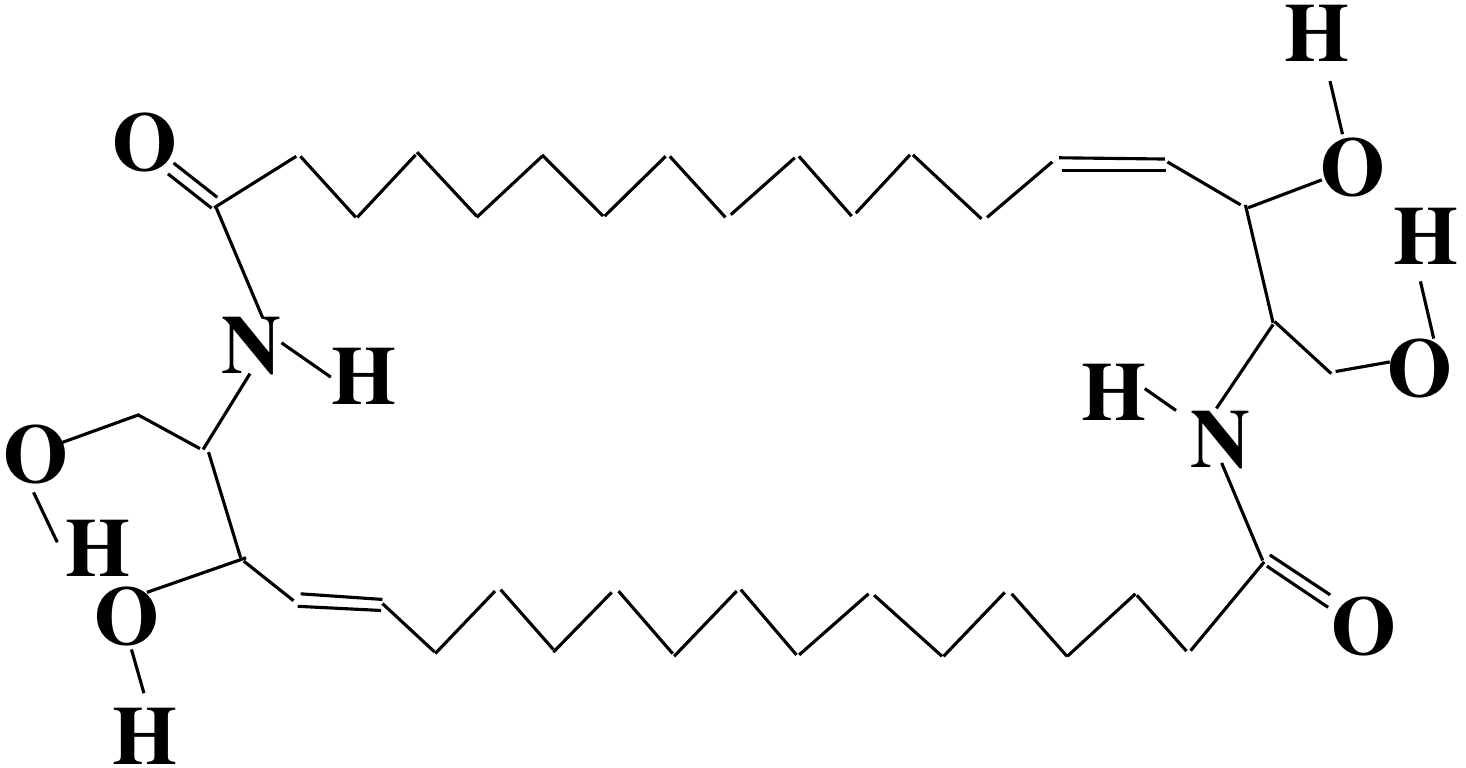}}
\centerline{(a)}
\includegraphics[width=0.95\linewidth,clip=]{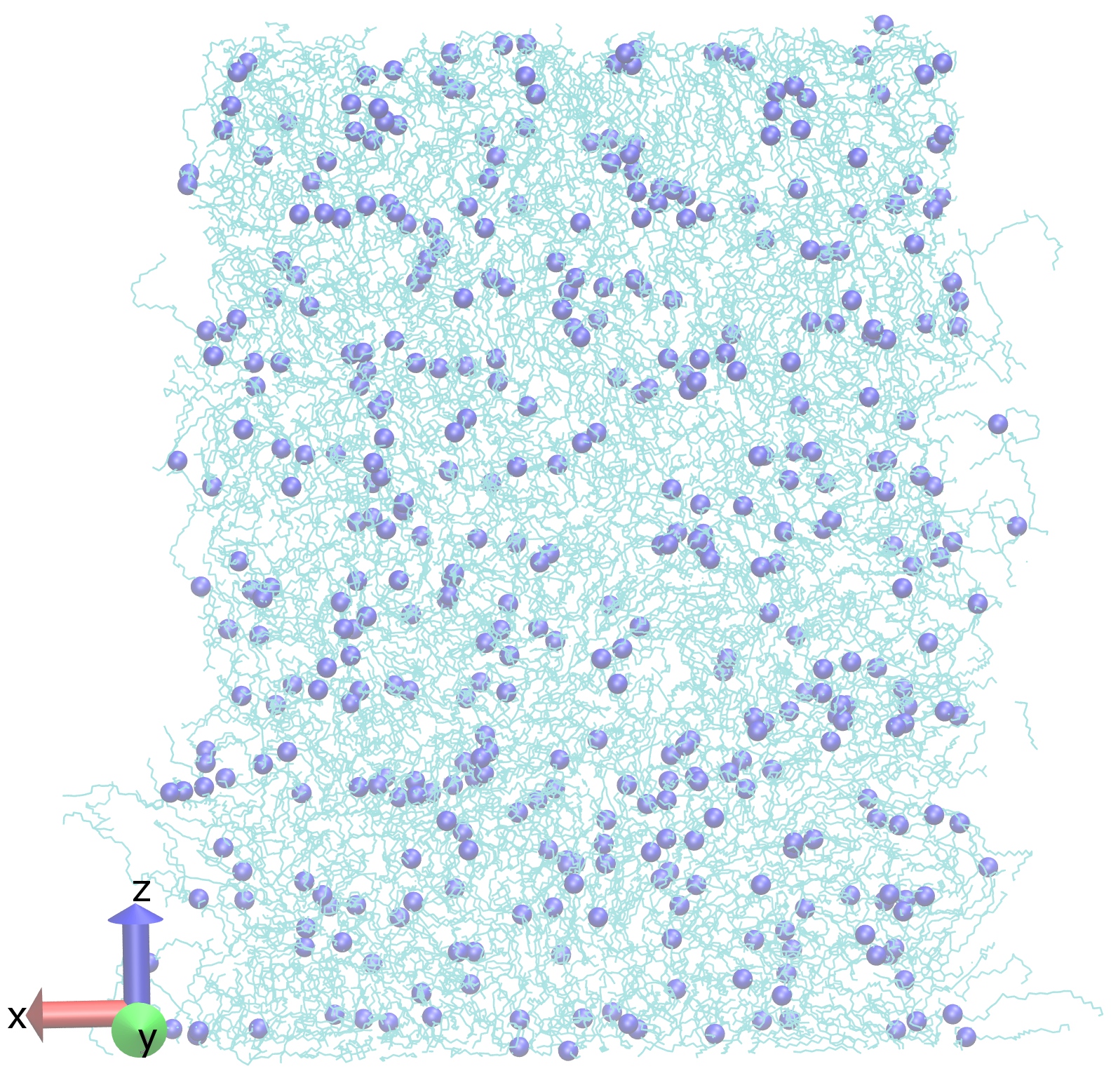}
\centerline{(b)} 
                             } \hfill
\parbox{0.49\linewidth}{
\centerline{\includegraphics[width=0.95\linewidth,clip=]{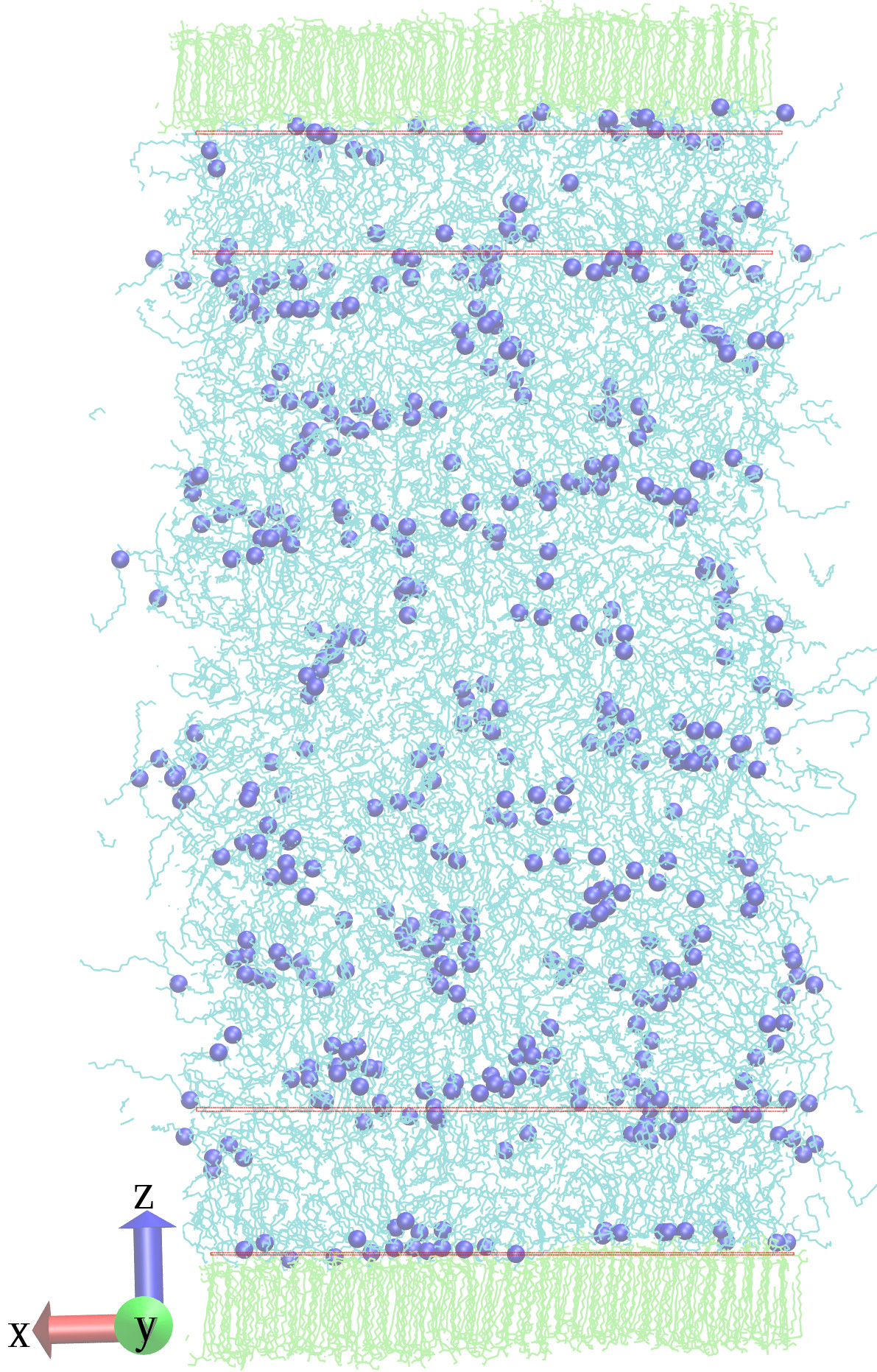}}
\centerline{(c)}
                       }}}
\caption{(a) Schematic representation of hypothetical CRW molecule
used to mimic corneocyte wall.
(b) Configuration of lipids (cyan) with continuous wall (not shown) along
the z-directions. Nitrogen atoms in the head groups are shown
as blue spheres. The  box size is 
$15.4\textrm{nm} \times 13.6\textrm{nm} \times 21.7\textrm{nm}$.
(c) Configuration of lipids  with CLE-mimic wall (lime)
constructed with CRW molecules show layering, in contrast to continuous wall
simulations. The two bilayers closest to the two walls are indicated
by drawn lines.  The  box size is 
$13.6\textrm{nm} \times 12.3\textrm{nm} \times 28.8\textrm{nm}$.
Only 3~nm slices along the y-direction are shown in (b) and (c). }
\label{fig.cle}
\end{figure}

{\bf Imposed layering near corneocyte:}
The corneocytes are surrounded by an envelope of 
grafted ceramide molecules whose fatty acid end is
believed to be covalently bonded to the protein network in the corneocyte
\cite{swartzendruber.CLE}
(Fig.~\ref{fig.sc.intro}a). The
highly hygroscopic corneocyte interior retains most of the water
in the SC \cite{bouwstra.corneocyte.03}, leaving an essentially dry lipid matrix
confined in the narrow space between corneocytes. The grafted
ceramides in the corneocyte-bound lipid envelope (CLE) presents
a patterned surface of favourable hydrogen-bond (H-bond) sites to this confined lipid matrix. 
By analogy with confined liquid crystals \cite{ajdari.91} and 
growth of colloidal crystals over patterned substrates \cite{hoogenboom.03},
we expect that a surface with such patterning of
favourable H-bonding sites will impose layering in the lipids.
We model the CLE as a wall of hypothetical 
molecules formed by joining the two tails of a symmetric CER~NS molecule with the
mirror image of its head group (Fig.~\ref{fig.cle}b). 
By construction, this molecule (termed CRW for corneocyte wall)
forms a stable flat layer structure. 

To be far from lamellar structure, we chose 
a final inverse-micellar phase 
configuration (2:2:1 lipid composition) 
as the starting point of this simulation. 
We partition the lipids along the $z$-direction by
freezing the water molecules 
in a narrow strip, making 
that region unfavourable for the lipid molecules \cite{supmat}.
The structure away from this frozen water layer remain
broadly unchanged. 
Next we slowly remove the water molecules, and equilibrate the lipids for
100ns between two continuous water walls.
No layering developed during these steps (Fig.~\ref{fig.cle}a). A layer of 
CRW was separately equilibrated with the
number of CRW chosen to match the lateral dimension of the lipid system.
The lipids were placed between the CRW walls.
Fig.~\ref{fig.cle}c shows the final configuration after subsequent
evolution at 340K for 1$\mu$s in the presence of the CRW wall. 
The lipid tails
predominantly align perpendicular to the wall. 
Close to the wall,
layering is near perfect and is parallel to the CRW walls. 
The ordering grows slowly due to low mobility
of the molecules.

{\bf Discussion:}
These simulations show that
the SC lipids form inverted micellar phases. A patterned wall mimicking 
the conrneocyte-bound lipid envelope induces layering in
SC lipids. A continuous wall failed to induce layering,
suggesting that the hydrogen-bonding with the wall molecules
plays an important role in forming multilayered structures.

These simulations  provide some rationale behind the synthesis pathway and the
lipid structures found in the SC {\em in~vivo}. 
The seemingly unnecessary step of adding and removing a sugar group \cite{Mizutani.09}
becomes meaningful by noting that the ceramide lipids would have formed
an inverse micellar phase if left without a large head group inside the cells in the SG, and thus
frustrate the vesicle trafficking mechanism necessary for their release. 
Once the sugar group is removed, if the molecules remain in the water-rich 
SG-SC boundary region they can form an inverse micellar phase,  leading
to the granular pattern seen in cryo-EM \cite{amoudi.cubic.cryoem.05}. 
These water-containing inverted structures may play a role in the
activity of anti-microbial peptides that are released along with
the lipid molecules \cite{aberg.amp.07} and possibly are
responsible for higher permeability in skin regions lacking corneocyte stacks
\cite{schatzlein.sc.confocal.98}. Such non-lamellar
structures have also been observed in some {\em in~vitro} depositions
\cite{plasencia.sc.confocal.07}. 
When transported 
into the region between corneocyte walls,
the patterned surface and confinement leads to a lamellar structure.
The growth of this lamellar structure is necessarily slow because
of low molecular mobility in the crowded confined environment. From
{\em in~vivo} measurement of recovery of permeability barrier, we
expect the time-scale of this patterned wall induced ordering to
be of order days \cite{grubauer.jlr.89} and the rate limiting step in the recovery process.
The polydispersity stops the lipids from forming
crystalline structures, thus providing some plasticity and
dissipation. The polydisperse tails also lower the defect
energies (as evidenced by the diffuse defect structure in fig.~\ref{fig.mult.unstable}d)
so that the lipids can envelope the corneocytes in three
dimensions without forming strongly localised defects and thereby 
compromising the barrier.

\begin{acknowledgments}
This work was supported by Yorkshire Forward through the grant
YFRID Award B/302 and part financed by the European Regional Development
Fund (ERDF). Computational resources were provided by 
SoftComp EU Network of Excellence. We gratefully acknowledge
helpful comments by Patrick Warren. 
CD thanks 
Lars Norl\'en for useful discussions and for sharing unpublished results. 
\end{acknowledgments}

\bibliographystyle{apsrev4-1}

%


\clearpage
\newpage
\setcounter{page}{1}
\setcounter{section}{0}
\setcounter{figure}{0}
\setcounter{equation}{0}
\leftline{\bf Supplementary material:}
\begin{flushleft}
Lamellar and inverse micellar structures of 
skin lipids: Effect of templating 
\end{flushleft}
\leftline{Chinmay Das, Massimo G. Noro and Peter D. Olmsted}
\hrule
\renewcommand{\thefigure}{S\arabic{figure}}
\renewcommand{\thepage}{S\arabic{page}}
\renewcommand{\thesection}{S\arabic{section}}
\renewcommand{\thesubsection}{S\arabic{section}.\arabic{subsection}}
\renewcommand{\thetable}{S\arabic{table}}
\renewcommand{\theequation}{S\arabic{equation}}

\section{Stratum Corneum  lipid molecules}
\begin{figure*}[htbp]
\centerline{\includegraphics[width=12cm,clip=]{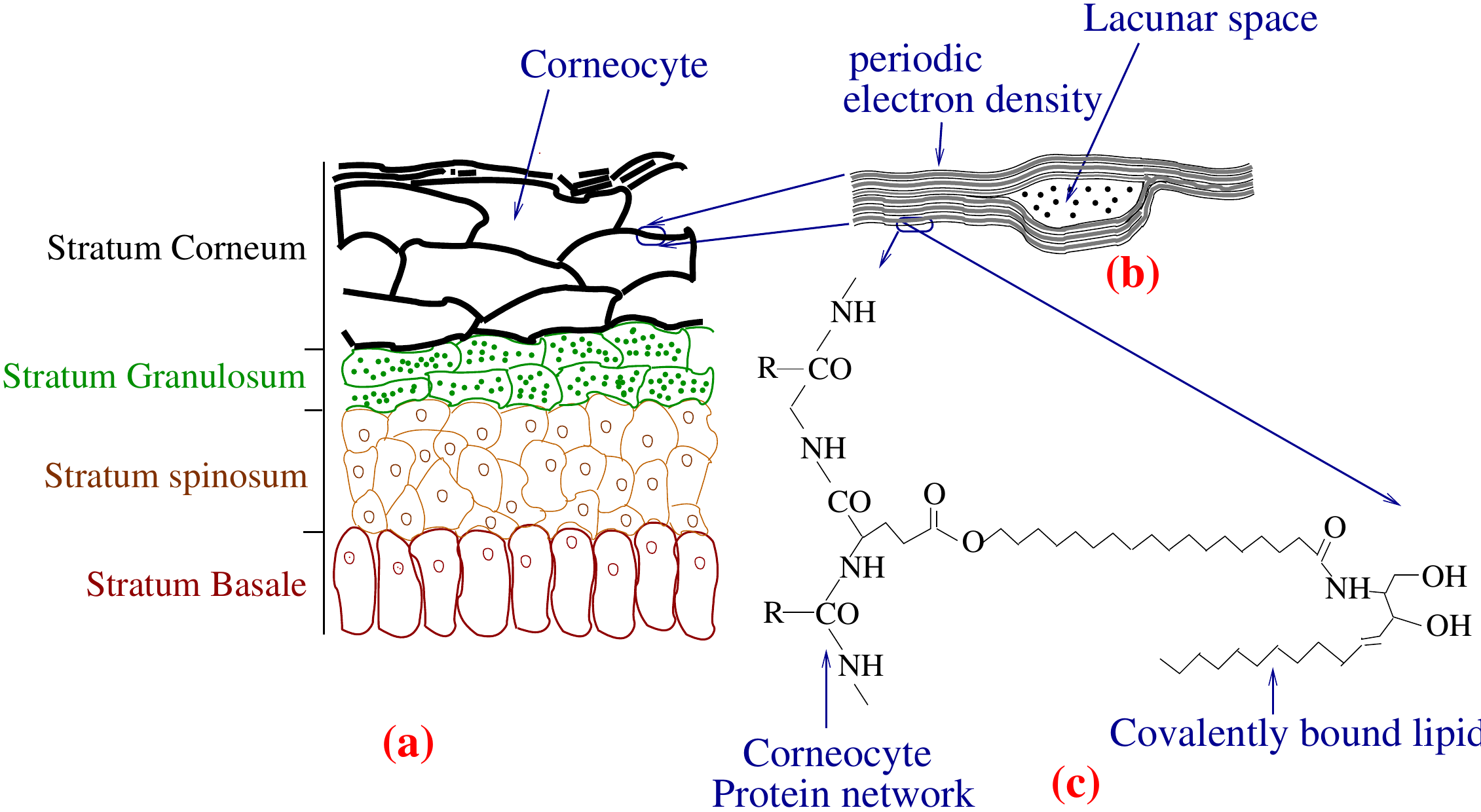}}
\caption{(a) Schematic representation of the layers in the epidermis. 
(b) The inter-corneocyte region is filled with specialized lipids 
whose electron density 
shows a multilamellar structure. The lamellar
regions are punctuated by {\em lacunar spaces} (schematic
figure based on [2,3]). 
(c) Some of the
lipids are covalently bound to the protein networks in the corneocytes
(schematic figure based on [4]). }
\label{fig.sc.schematic}
\end{figure*}

Epidermis, the outer layer of the skin, has a stratified organization and
is named according to the visual appearance under optical microscope
(Fig.~\ref{fig.sc.schematic}.a). The {\em stratum
corneum} (SC) is the outermost layer of the epidermis, comprising dead 
pancake-like flattened cells (corneocytes) in a lipid matrix. 
Keratinocytes,
the dominant cell type in the {\em stratum basale}, the innermost
layer of epidermis, migrate through the intervening layers while changing
in shape and content. In the {\em stratum granulosum}, immediately 
below the SC,
keratinocytes over-produce keratin and secrete specialized lipids. 
By the time keratinocytes migrate to SC, the neuclei and other internal
organelles disintegrate and the cell attains a comparatively rigid structure
through keratin network. The dead keratinocytes are called corneocytes. 
The lipid matrix in the SC provides the main barrier against water
loss and invasion of foreign pathogens [1].

The lipid matrix in the extracellular space of the corneocytes 
shows a  multilamellar structure in electron microscopy 
(Fig.~\ref{fig.sc.schematic}.b) [2]. 
This multilamellar structure is often punctuated by 
{\em lacunar spaces} where the periodic structure in electron density
is lost. The lipids adjacent to the corenocytes are covalently
bound to the protein network of the corneocytes forming 
the {\em corneocyte-bound lipid envelope} (CLE) (Fig.~\ref{fig.sc.schematic}.c),
 which shows up as  electron-dense regions in electron microscopy.

\begin{figure*}[htbp]
\centerline{\includegraphics[width=12cm,clip=]{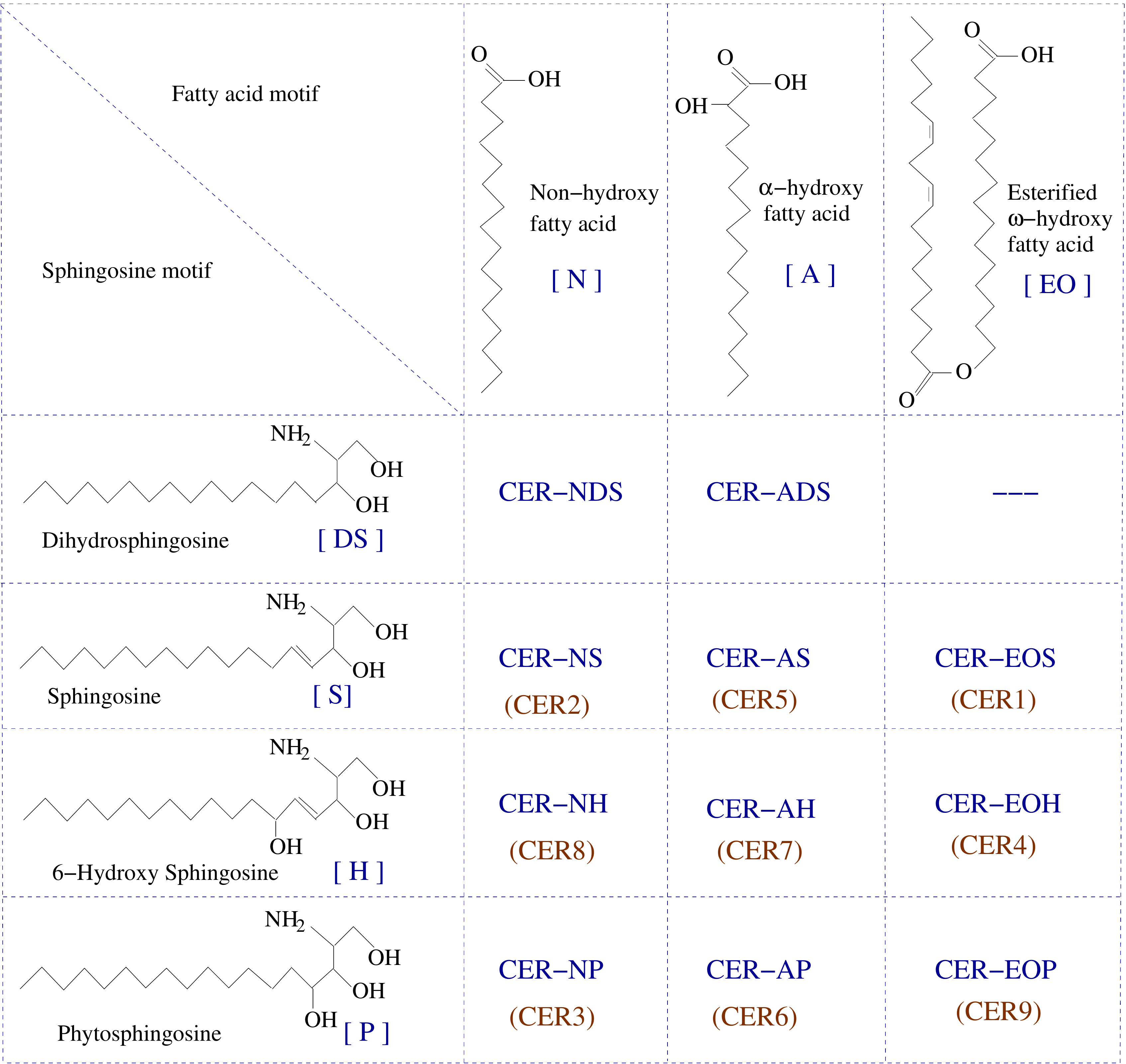}}
\caption{Ceramide family in the SC lipid matrix. Each of the members
consists of a sphingosine motif bonded via a peptide (amide) bond to 
a fatty acid motif. The
fatty acid motif shown here is palmitic acid (16 carbon). Ceramides
in the SC contain variable length fatty acids (typically 20 to 30
carbon atom long). (Adapted from [5]) }
\label{fig.cer.family}
\end{figure*}

The main components of the SC lipid matrix are the ceramide sphingolipids (CER),
cholesterol (CHOL) and free fatty acid (FFA). There are 11 families of CER
molecules found in the SC (Fig.~\ref{fig.cer.family}), differing  by the numbers
and the positions of the hydroxyl groups. Each member of these CER families
shows additional polydispersity in the number of carbons
in the fatty acid tail motif, leading to more than 300
different CER molecules [6]. We refer to the molecules
by their family names followed by the number of carbon atoms
in the fatty acid motif. Thus, CER~NS~24 refers to CER~NS with
lignoceric acid as the fatty acid motif. 
Similarly the different FFA molecules are distinguished by
the number of carbon atoms. In the literature, ceramide familes
often are distinguished by the order of their appearance in
the chromatograph: CER~EOS is often refereed to as ceramide 1,
CER~NS as ceramide 2 and CER~NP as ceramide 3.

\section{Simulation Details}
All simulations were carried out with the molecular dynamics software GROMACS
[7,8]
with Nos\'e-Hoover thermostats separately coupled
to the lipid molecules and the water molecules, and with
a Parrinello-Rahman barostat.
Bond lengths were constrained with the LINCS algorithm
for the lipid molecules and with the SETTLE algorithm
for water molecules.
The time steps for MD simulations were 2~fs.

\begin{figure*}[htbp]
\centerline{\includegraphics[width=12cm, clip=]{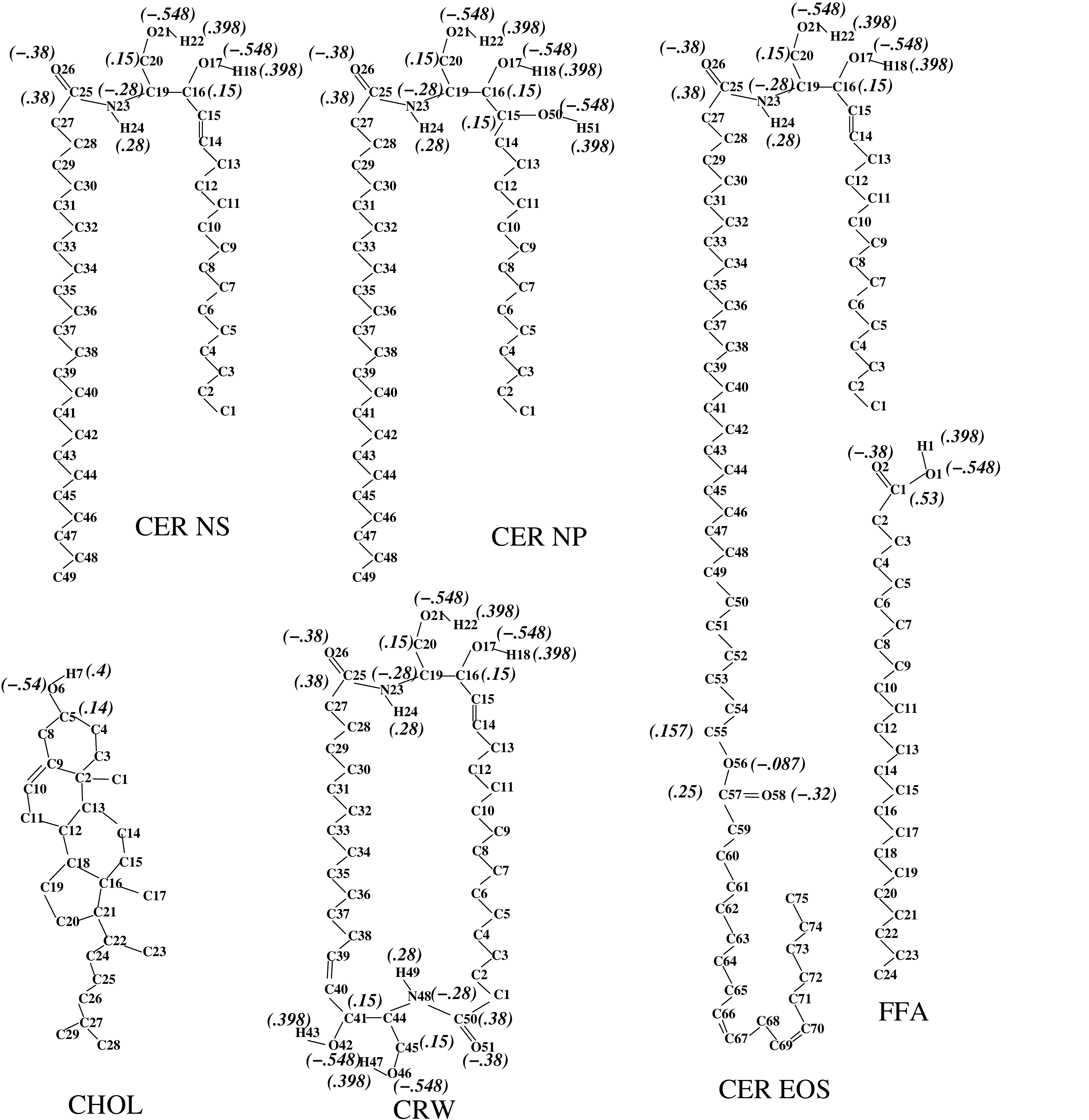}}
\caption{Partial charges used for the different lipid molecules.}
\label{fig.sup.charge}
\end{figure*}

We used the `Berger' force-field [9] for the lipid molecules, 
which accurately reproduces experimental results 
for different lipid systems [10].
We have used the same force-field in the past for
simulations of a smaller subset of SC lipid molecules for studying small
fully hydrated bilayers and water permeation [11,12]. 
Water is modeled with
the SPC potential [13]. The partial charges used are
shown in Fig.\ref{fig.sup.charge}.

We use a group-based cut-off with the cut-off length $1.2\,\textrm{nm}$ for
both the van~der Waals and electrostatic interactions. In our previous 
simulations [11], we have explicitly compared simulations 
involving a group-based cut-off 
with particle-mesh Ewald summation (PME) simulations. The effect
of the long range electrostatics was found to be negligible. 
In simulations of lipid bilayers, long range electrostatics often
plays an important role. The reasons for the negligible contribution
of long range electrostatics in our simulations are two-fold: Firstly, 
the SC lipids are uncharged at skin pH, with only  small partial charges 
contributing to the electrostatic interaction. Thus, the
dipole moments involved are about $15$ times smaller than the
typical phospholipid dipole created from well separated N$^{-}$ and
P$^{+}$ charges. Secondly 
 phospholipid dipoles have a component normal 
to the water-lipid interface, which can give 
rise to a macroscopic dipole moment.  By contrast, the small dipoles
from the partial charges in the SC lipid system, even in a 
bilayer arrangement, orient themselves in random directions. Thus, the
effect of the electrostatic interaction beyond the cut-off becomes much smaller
than the thermal energy.

\begin{table}[htbp]
\centering
\begin{tabular}{|c|l|l|l|l|}
\hline \hline
molecule & fatty acid tail & \multicolumn{3}{|c|}{Number of molecules}\\ \cline{3-5}
family & (number of carbons) & setA & set B & set C   \\\hline \hline
\multicolumn{1}{|c|}{\multirow{4}{*}{CER~EOS}} &
\multicolumn{1}{|c|}{30} & 30 & 30 & 20  \\ \cline{2-5}
\multicolumn{1}{|c|}{}& 
\multicolumn{1}{|c|}{32} & 80 & 80 & 52  \\ \cline{2-5}
\multicolumn{1}{|c|}{}& 
\multicolumn{1}{|c|}{33} & 30 & 30 & 20  \\ \cline{2-5}
\multicolumn{1}{|c|}{}& 
\multicolumn{1}{|c|}{34} & 60 & 60 & 40  \\ \hline \hline
\multicolumn{1}{|c|}{\multirow{6}{*}{CER~NS}} &
\multicolumn{1}{|c|}{22} & 100 & 100 & 68  \\ \cline{2-5}
\multicolumn{1}{|c|}{}& 
\multicolumn{1}{|c|}{24} & 200 & 200 & 132  \\ \cline{2-5}
\multicolumn{1}{|c|}{}& 
\multicolumn{1}{|c|}{25} & 100 & 100 & 68 \\ \cline{2-5}
\multicolumn{1}{|c|}{}& 
\multicolumn{1}{|c|}{26} & 300 & 300 & 200 \\ \cline{2-5}
\multicolumn{1}{|c|}{}& 
\multicolumn{1}{|c|}{28} & 200 & 200 & 132 \\ \cline{2-5}
\multicolumn{1}{|c|}{}& 
\multicolumn{1}{|c|}{30} & 100 & 100 & 68  \\ \hline \hline
\multicolumn{1}{|c|}{\multirow{5}{*}{CER~NP}} &
\multicolumn{1}{|c|}{24} & 80 & 80 & 52  \\ \cline{2-5}
\multicolumn{1}{|c|}{}& 
\multicolumn{1}{|c|}{26} & 136 & 136 & 92  \\ \cline{2-5}
\multicolumn{1}{|c|}{}& 
\multicolumn{1}{|c|}{28} & 208 & 208 & 140  \\ \cline{2-5}
\multicolumn{1}{|c|}{}& 
\multicolumn{1}{|c|}{30} & 240 & 240 & 160  \\ \cline{2-5}
\multicolumn{1}{|c|}{}& 
\multicolumn{1}{|c|}{32} & 136 & 136 & 92  \\ \hline \hline
\multicolumn{1}{|c|}{\multirow{7}{*}{FFA}} &
\multicolumn{1}{|c|}{20} & 100 & 50 & 64  \\ \cline{2-5}
\multicolumn{1}{|c|}{}& 
\multicolumn{1}{|c|}{22} & 200 & 100 & 136  \\ \cline{2-5}
\multicolumn{1}{|c|}{}& 
\multicolumn{1}{|c|}{24} & 800 & 400 & 536  \\ \cline{2-5}
\multicolumn{1}{|c|}{}& 
\multicolumn{1}{|c|}{25} & 200 & 100 & 136 \\ \cline{2-5}
\multicolumn{1}{|c|}{}& 
\multicolumn{1}{|c|}{26} & 460 & 230 & 304 \\ \cline{2-5}
\multicolumn{1}{|c|}{}& 
\multicolumn{1}{|c|}{28} & 200 & 100 & 136  \\ \cline{2-5}
\multicolumn{1}{|c|}{}& 
\multicolumn{1}{|c|}{30} & 40 & 20 & 24  \\ \hline \hline
CHOL & \multicolumn{1}{|c|}{--} & 2000 &2000 &1332 \\ \hline \hline
SOL & \multicolumn{1}{|c|}{--} & 50000 & 43600 &20000 \\ \hline \hline
\multicolumn{2}{|l|}{Molar ratio} &1:1:1 & 2:2:1 & 1:1:1 \\
\multicolumn{2}{|l|}{CER:CHOL:FFA} & & &  \\ \hline \hline
\end{tabular}
\caption{Numbers of molecules used in the simulations. For the sets A and B, we use three different initial
configurations. The three configurations from set A are referred to as sets Aa, Ab and Ac.  A similar suffix is
used for set B.} 
\label{tab.molnum}
\end{table}


Table~\ref{tab.molnum} gives the number of molecules used in the
different simulations. Guided by the mass-spectroscopic profiles
for the ceramides [14,15], 
we choose a fixed molar ratio 1:5:4 for the CER~EOS : CER~NS : CER~NP. 
In doing so, we assume that the ceramides with esterified $\omega$-hydroxy
fatty acids (CER~EOS, CER~EOH and CER~EOP) behave similarly and
CER~EOS alone represents all three of these families in our simulations.
Similarly, ceramides that do not have such esterified fatty acid
and have a double bond in the sphingosine motif are represented by
CER~NS. CER~NP accounts for rest of the ceramides (all of which
do not have double bonds in the sphingosine motif). 
Set~A and set~C consider CER, CHOL and
FFA in 1:1:1 molar ratio. Set~B considers a molar ratio of
CER : CHOL : FFA = 2:2:1. Set~A and set~B contains 30 wt\% 
water. In both of these two sets (A and B), three separate simulations
were performed with different initial conditions, which we will distinguish as set~Aa, set~Ab and 
so on. Set~C was used to look at the stability of pre-formed multilayer
arrangements (Details follow in S2.4).

\subsection{Inverse micellar phase}
Configurations with a single molecule of each of the lipid species were 
energy minimized in vacuum separately. For the simulations
in sets A and B, a large simulation box
was divided into small grids with dimensions large enough to accommodate
the longest lipid species. The desired number of 
energy-minimized lipid molecules were placed with random orientations
in randomly selected unoccupied grid points. 
We maintain 
a finer grid which notes occupied sites by individual lipid atoms. 
The required number of water molecules were placed with random orientations 
at randomly chosen unoccupied sites of the fine grid. 

The configurations were energy minimized and evolved with NPT MD steps
using a Nose-Hoover thermostat and a Parrinello-Rahman
barostat for $50\,\textrm{ns}$ at 340~K. Standard periodic
boundary conditions in a cubic simulation box and 
isotropic pressure coupling were used in all cases except for
set Bc.

\begin{figure*}[htbp]
\centerline{\includegraphics[width=12cm,clip=]{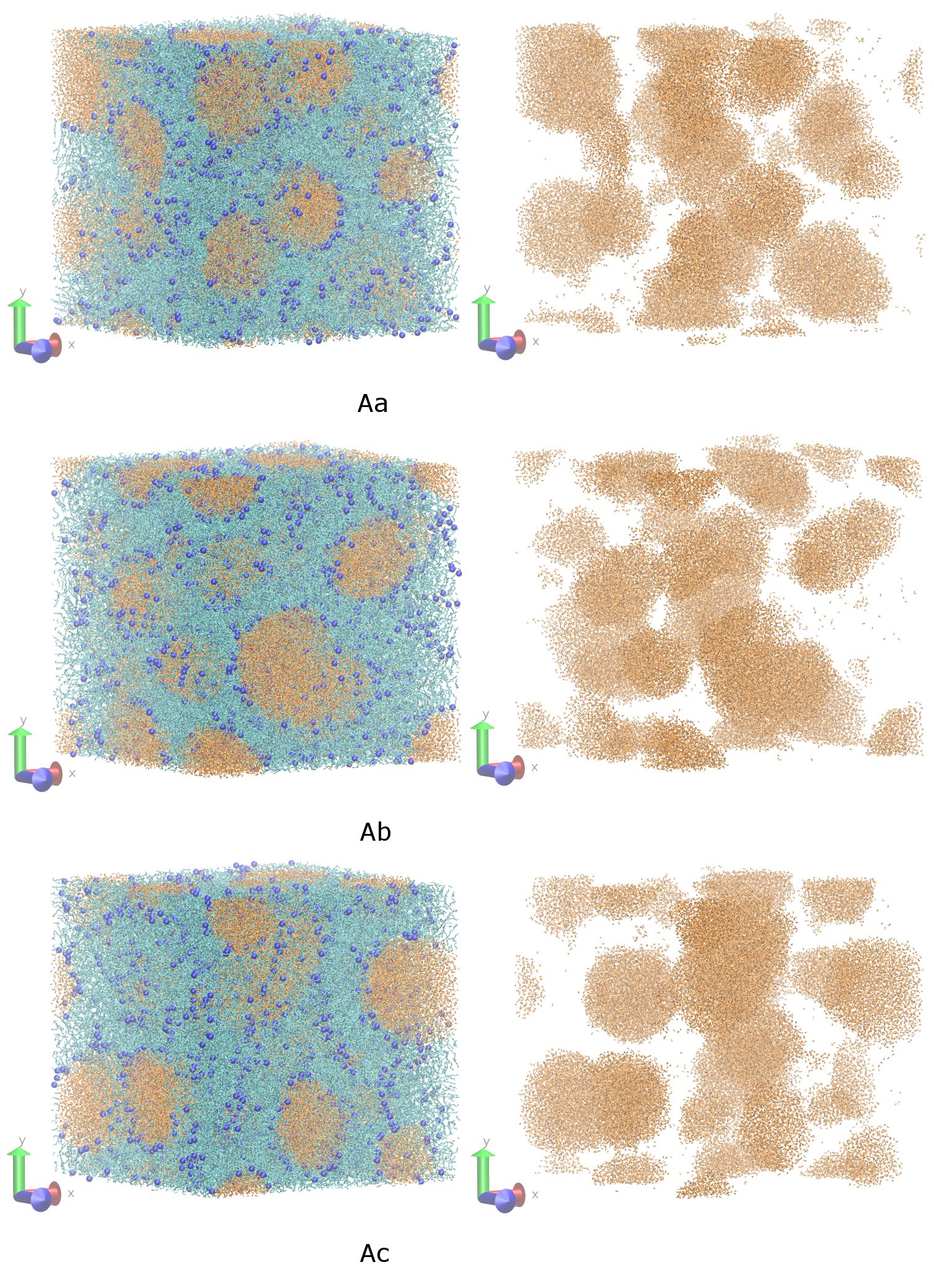}}
\caption{Final configurations of the three simulations with 1:1:1 
molar concentration of CER, CHOL and FFA (Sets Aa, Ab and Ac). 
The right-hand panel shows 
only the water molecules to highlight the disconnected water
clusters. The box dimension is $\sim 18.8 \;\; \textrm{nm}$ in all
three configurations.}
\label{fig.seta.fin}
\end{figure*}

\begin{figure*}[htbp]
\centerline{\includegraphics[width=12cm,clip=]{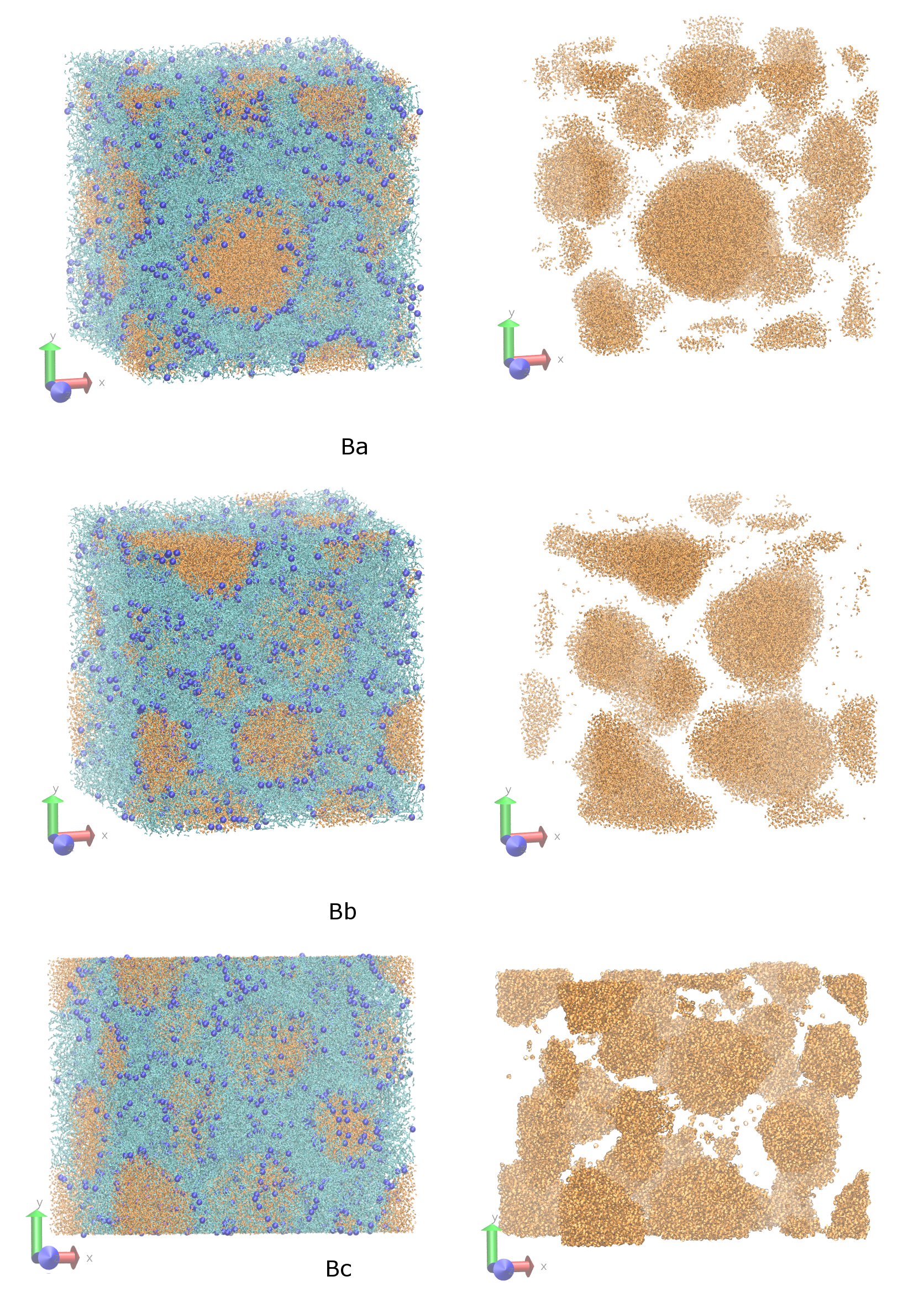}}
\caption{Final configurations of the three simulations with 2:2:1 
molar concentration of CER, CHOL and FFA. The right-hand panel shows 
only the water molecules to highlight the disconnected water
clusters. Sets Ba and Bb are 50~ns after density equilibration.
Set Bc is evolved for a further 50~ns at 340K and another 100~ns at 300K.
Sets Ba and Bb are in cubic box with dimension $\sim$ 17.9~nm.
The box size for set Bc is 20.3~nm $\times$
17.1~nm $\times$
16.2~nm.} 
\label{fig.setb.fin}
\end{figure*}

For one of the 2:2:1 
composition system (set~Bc), we selected a rectangular box with 
$x-y$ dimensions nearly four times larger than the $z$ direction 
($L_x = L_y = 175~nm$, $L_z=56~nm$). Once the initial `gas'-phase system 
reached atmospheric pressure, 
an anisotropic barostat with a large (1000 bar)
pressure applied along the $x-y$ directions  and atmospheric
pressure applied in the $z$ direction was used to achieve a nearly
cubic box. The resulting deformation aligns the long molecules preferentially
along the $z$ direction. After this, the pressure couplings were set to
atmospheric pressure in all three directions and the system was evolved 
for $50\,\textrm{ns}$ at 340~K. The initial alignment did not affect the
transition to the final inverse micellar arrangement with on average 
isotropic lipid 
orientations. 
We evolved this set~Bc for a further 50~ns at an elevated 
temperature
of 350~K and then finally for another 100~ns at 300~K. The inverse
micellar arrangement was found to be robust under change in temperature.
The final configurations of the different sets are presented in 
Fig.~\ref{fig.seta.fin} and Fig.~\ref{fig.setb.fin}.

\subsection{Surface-templated lipid system}

\begin{figure*}[htbp]
\centerline{\includegraphics[width=12cm,clip=]{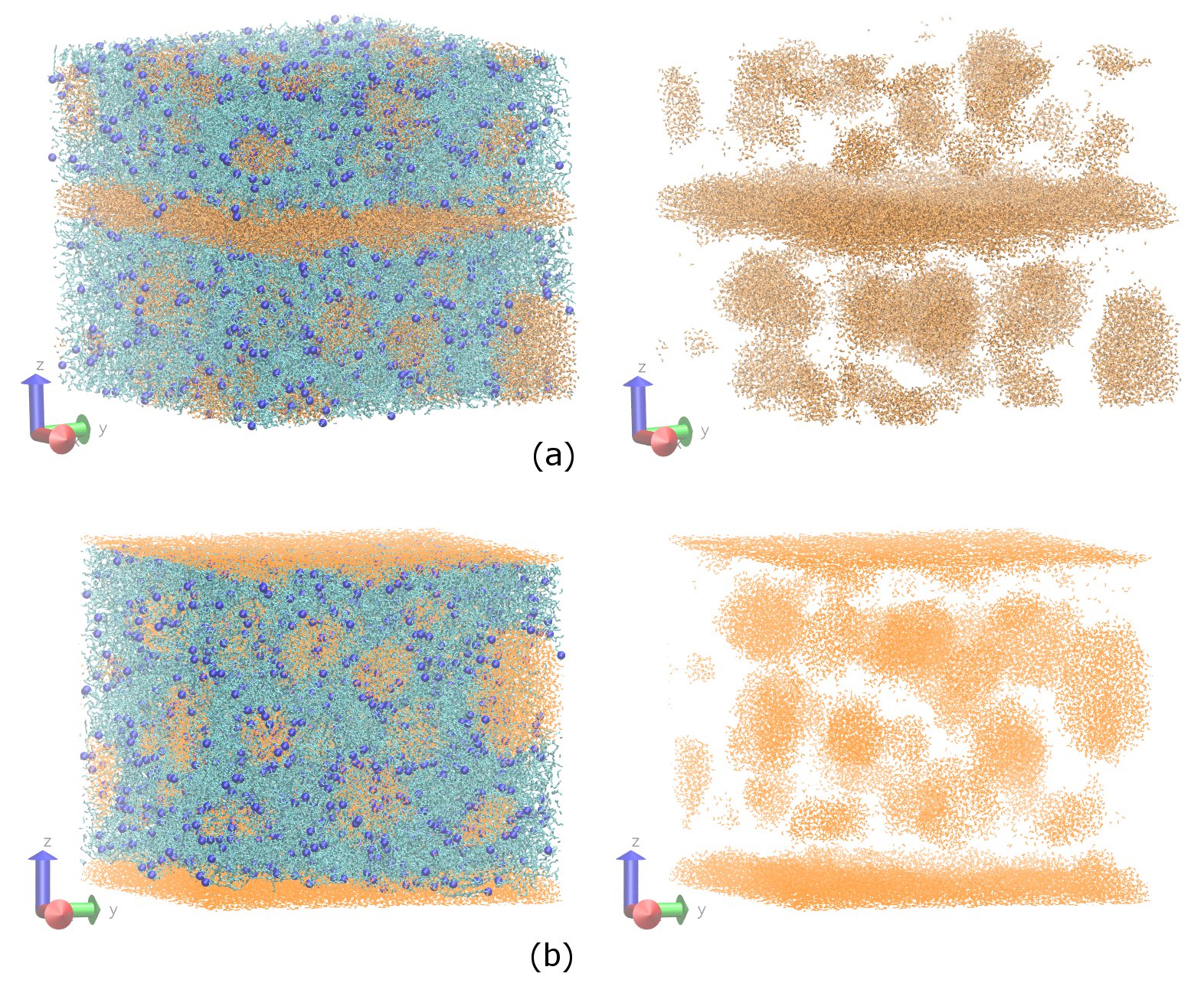}}
\caption{(a) Simulation of set Bc with water molecules frozen
in $1\,\textrm{nm}$ region at the center of the box. Box size is
19.5~nm $\times$ 17.4~nm $\times$ 16.3~nm. (b)
The procedure allows us to introduce a wall keeping the inverse
micellar arrangement. The box size is 19.5~nm $\times$ 17.4~nm $\times$
16.7~nm.}
\label{fig.fzsol}
\end{figure*}

In the final inverse micellar configuration of set~Bc 
we selected a $1\,\textrm{nm}$ zone, within which the water molecules were 
frozen along the $z$-direction.
We repeatedly carried out short ($1\,\textrm{ps}$) NVT simulations
and identified any new water molecules drifting in the selected zone. These
water molecules in turn were added to the frozen list. Locally this
creates enough perturbation to move the lipid molecules
out of this zone.
Only a few lipid tails stay stuck in the frozen 
water layer. They were pulled out of the
water layer by moving them at a constant speed (0.01~nm/ps)
away from the water layer. The configuration created by this process
is shown in Fig.~\ref{fig.fzsol}~a. The inverse micellar structure
away from the frozen water layer remain unchanged during further
evolution (50~ns, Fig.~\ref{fig.fzsol}~b). 

\begin{figure*}[htbp]
\centerline{\includegraphics[width=12cm,clip=]{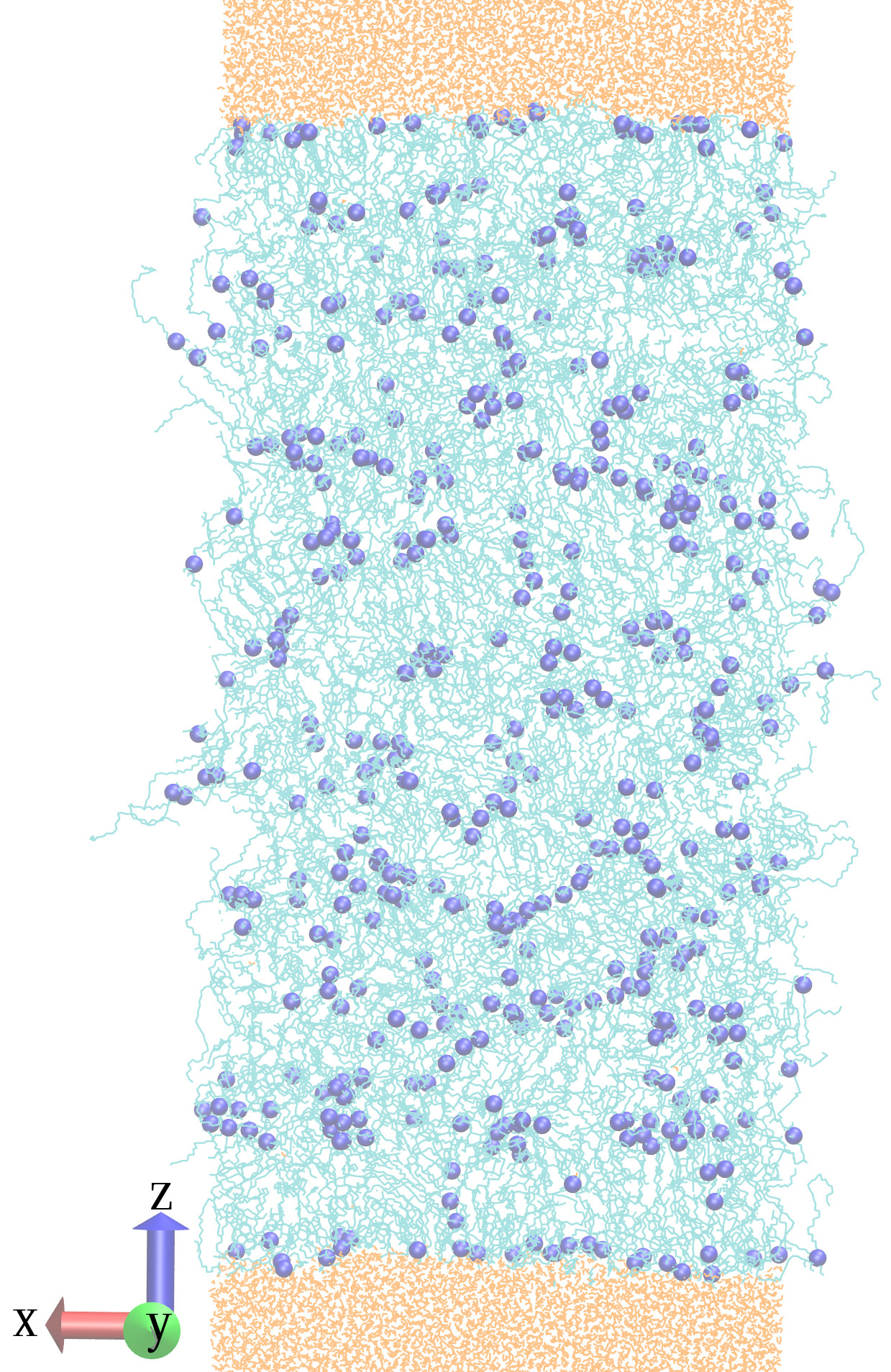}}
\caption{ Lipids from set Bc after evolving for 100~ns, placed between 
water walls along the $z-$direction. Box dimensions are
13.4~nm $\times$ 12.1~nm $\times$ 32.1~nm. Only a 3~nm slice along
the y-direction is shown.}
\label{fig.discwall}
\end{figure*} 

The frozen water layer allows us to introduce a wall along
the $z$-direction mimicking a continuous distribution of
water molecules.  The walls interact with the atoms as an integrated 
Lennard-Jones (9-3) 
potential with interaction parameters chosen to be that of atom type
OW (oxygen of water) at density 33.44 atoms/nm$^3$. After 50~ns
of NPT simulation, no qualitative change in the lipid
arrangement was observed (Fig.~\ref{fig.fzsol}~b).

In SC, the lipids between highly hygroscopic corneocytes are
essentially free of water [16]. To simulate at conditions similar
to that between corneocytes, 
we {\em dry} the lipid structure by slowly removing  the
inner water molecules
(10 randomly water molecules were removed every 100~ps of NPT simulation), 
and evolve the system for another 50~ns
with a continuous water wall. 
The anisotropy introduced by the walls along the $z$-direction
 changes the box dimension
significantly (Fig.~4a). But no signature of
layering was observed. Specifically, the hydrocarbon tails were
found to be equally likely to be parallel to the wall as perpendicular to
the wall for lipids close to the boundaries.

In the SC, a layer of ceramide
molecules are covalently bonded to the proteins in corneocyte. Thus
the CLE surface  presents itself as a
well aligned layer of ceramide head groups to the stratum corneum
lipids. In our next step we try to mimic this surface. 

First we replace the continuous wall with a layer of 
SPC water molecules (24000 molecules) and 
use 100~ns of NPT simulation. Discrete water molecules
provide transient hydrogen bonds that align the lipids closest to the
boundary, and the lipid tails align predominantly along the $z$-direction
(Fig.~\ref{fig.discwall}). 
From a separate simulation of a hydrated bilayer comprising CER~NS~24:0, a 
single molecule with a hairpin conformation was isolated. The head group
(atoms between C16 till O26) was duplicated (as atoms C41 till O51).
The new copy of the head group was mirrored first about $z$-axis and then
about the nitrogen atom. Finally the new group was given a rigid translation
by identifying the terminal CH3 group of the fatty acid chain of original
molecule as C41 of the duplicated head group. We term this artificial
lipid molecule CRW. Because of the symmetry of this molecule, it 
has  zero curvature and a flat layer structure will be
its preferred arrangement.
All interaction parameters of the new second head group are identical to 
the first head group. The molecule was placed in vacuum and energy minimized.

The existing lipid box in which we wanted to include the CRW wall had $x-y$
box dimensions of 13.37~nm $\times$ 12.10~nm. Guided by the area/lipid of
CER~2 bilayers [11], we placed 408 CRW 
molecules in a roughly triangular lattice with the $x-y$ dimensions of 
the box identical to the SC lipid system. 
We place two continuous water walls along the $z$-direction at separation of 3.8~nm.
The molecules were
energy minimized, and then equilibrated over 0.5~ns with NVT MD simulations at
340K.

This equilibrated CLW layer was included in place of the
water wall in the SC lipid box 
The configuration was energy minimized and evolved for further 1~$\mu$s 
at 340~K with NPT simulation. The final configuration of this
system is shown in the main
body of the paper (Fig.~4c). Close to the two walls,
the figure shows aligned bilayers. Further away from the walls, 
patches of bilayer can be seen that are not oriented with the wall.
The long timescale of realignment of the interior bilayer patches
may explain the long delay between the lipid release from SG and
recovery of the permeability barrier.

\subsection{Signature of inverse micellar phase in hydrated bilayer}
CER~NS tails are identical to sphingomyelin (SM) except for the absence of
the phosphocholine head group. The absence of the phosphocholine headgroup
leads to an area/lipid for the CER~NS molecules
 in a bilayer arrangement [11] 
that is only around 70\% of that for SM in a hydrated bilayer
[17]. Since SM
forms a stable bilayer, from geometric consideration alone 
[18], one would 
expect CER to form an inverted phase.
In fact, in all but one of the several crystal structures 
CER adopts a splayed
chain arrangement [19] to release the packing 
frustration of accommodating
a small head and comparatively bulky tails.

While a small bilayer with full hydration is indefinitely stable in simulations,
the local stress tensor across the bilayer 
contains information about the spontaneous curvature
preferred by the molecules. The integral of the first moment of the 
difference in the lateral ($P_{LAT}$) and the normal 
pressure  ($P_{zz}$) profile is related to the spontaneous curvature
$c_0$ and the bending elastic constant $\kappa$ for 
(half of) the bilayer through [20]
\begin{equation}
\kappa c_0 =  2 \int z \left[P_{LAT} (z) - P_{zz} (z)\right] dz.
\label{eq.latp}
\end{equation}
Here, $P_{LAT} \equiv \frac{1}{2} \left[P_{xx} + P_{yy}\right]$ and
the integral is assumed to be over a single monolayer. 
In the literature, this expression is often written in terms of 
tension, which is the negative of the lateral pressure. Also, different
prefactors are used for different definitions of the bending modulus.

\begin{figure*}[htbp]
\centerline{\includegraphics[width=12cm, clip=]{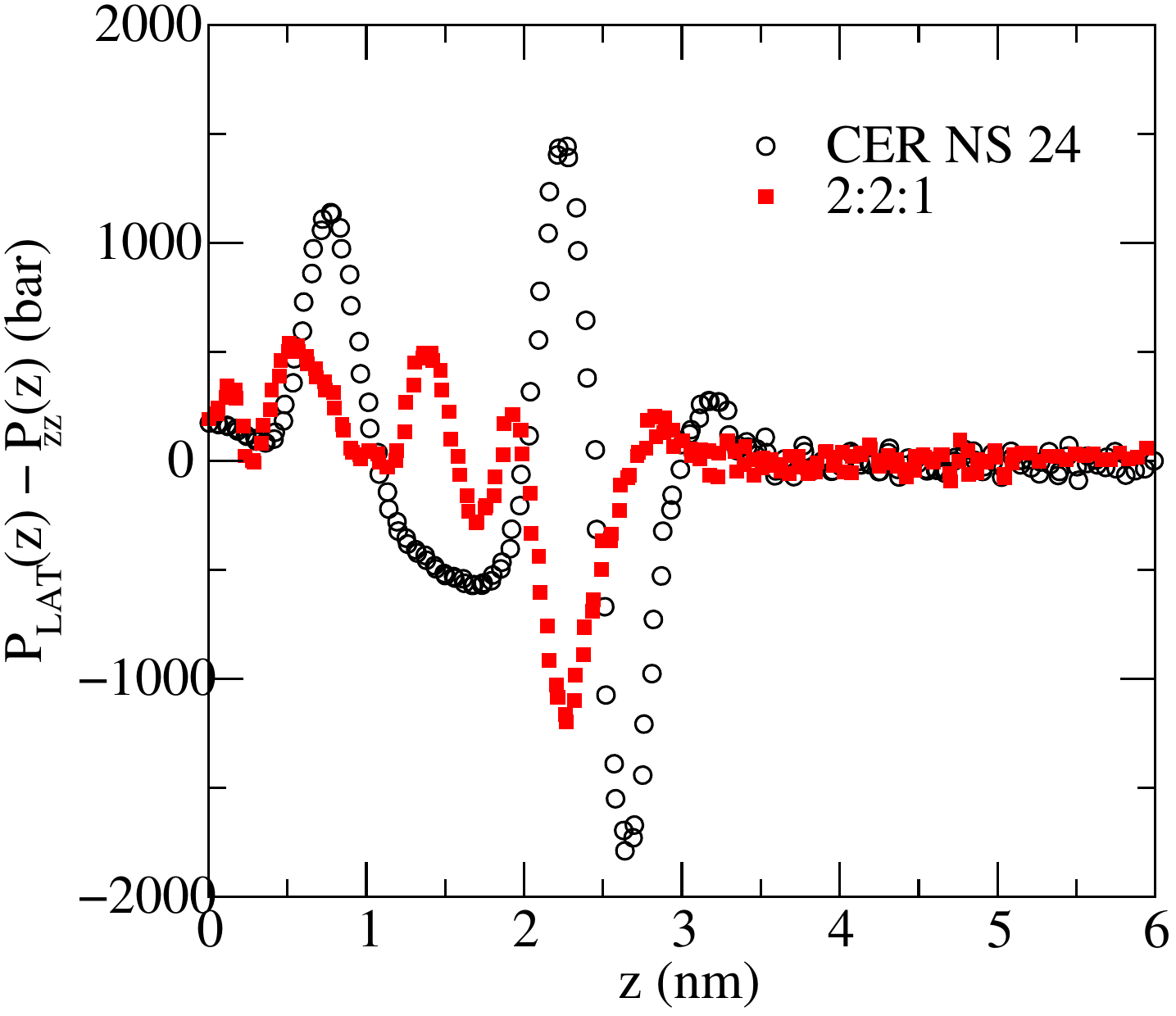}}
\caption{Lateral pressure profile for CER~NS~24 bilayer (open circles) and
bilayers of CER~NS~24, FFA~24 and CHOL at 2:2:1 molar ratio (squares).
The distance $z$ is measured from the bilayer center of mass and
the data is averaged over both the monolayers. 
(Adapted from [11])}
\label{fig.sup.localp}
\end{figure*}
Calculation of the local microscopic pressure tensor is costly and
we have not attempted such calculations for these 
simulations. However, we have re-analyzed the result of a CER~NS~24 hydrated
bilayer and  a 2:2:1  hydrated bilayer comprising  CER~NS~24, FFA~24 and CHOL
from [11]. In Fig~\ref{fig.sup.localp} we show the lateral pressure profiles
at 340K
after smoothing the data by combining results
from both the monolayers because of the up-down 
symmetry, and a further three point smoothing.

Using the bending modulus values from [11] 
($6.6\times 10^{-11} \; \textrm{erg}$
for the CER~NS bilayer and $3.3 \times 10^{-11} \; \textrm{erg}$ for the
2:2:1 bilayer), numerical integration of eq.~\ref{eq.latp}
then gives $1/c_0 \sim - 44 \;\textrm{nm}$ for (half of) the CER~NS bilayer and
$1/c_0 \sim - 15 \;\textrm{nm}$ for (half of) the mixed lipid bilayer. The sign of
the curvature points towards equilibrium inverse phases and the magnitude
is similar to the length-scale of the observed checkerboard 
pattern in cryo-EM 
[2].

\subsection{Pre-formed multilayers}
Separate leaflets with the same number of lipid molecules 
in each leaflet were prepared by randomly placing
the lipid molecules with the CER tails in hairpin arrangements 
and with the
head groups at the same $z$. Four such leaflets were energy
minimized and joined to form a double bilayer structure with sufficient 
separation between the leaflets to accommodate the longest lipid species 
without overlap with the adjacent leaflet. Continuous water wall was
placed along the $z$-directions.
The structure was repeatedly
compressed in $x-y$ direction by 1\%, 
the leaflets were moved closer to each other 
by 0.001~nm, and energy minimized. This ensures that the lipids remain
in multilayer arrangement, while locally deforming to allow a liquid-like
dense configuration. Once the internal pressure reached one atmosphere,
a layer of water (20000 molecules, $\sim 1~\textrm{nm}$) 
was placed in place of the wall, and for 1~ns only the
water molecules were evolved with NVT simulation while keeping the lipid
molecules frozen. Finally the constraints were removed and NPT steps
were used to further evolve the system for 260~ns.
We also simulated multilayers with identical preparation
steps except with a thicker water layer (25000 molecules).
Multilayers remain stable for the entire simulation duration
(260~ns) in such cases.

\subsection{Preparation of the simulation snapshots}
The figures showing simulation snapshots were prepared with
the package Visual Molecular Dynamics (VMD, \url{www.ks.uiuc.edu/Research/vmd}),
rendered with Persistence of Vision Raytracer 
(POV-Ray, \url{www.povray.org}) and annotated with 
GNU image Manipulation Program (GIMP, \url{www.gimp.org}). 

\section*{References}
\setlength{\parindent}{0pt}
\setlength{\parskip}{6pt}
[1] Ruth~K. Freinkel and David~T. Woodley, eds.
 {\em The biology of the skin}.
 (Parthenon Publishing, London, 2001).

[2] A.~Al-Amoudi, J.~Dubochet, and L.~Norl\'en,
 {\em J. Invest. Derm.}, {\bf 124}, 764 (2005).

[3] S.~Paliwal, G.K. Menon, and S.~Mitragotri,
 {\em J. Invest. Derm.}, {\bf 126}, 1095 (2006).

[4] D.C. Swartzendruber, P.W. Wertz, K.C. Madison, and D.T. Downing,
 {\em J. Invest. Derm.}, {\bf 88}, 709 (1987).

[5] J.~Ishikawa, H.~Narita, N.~Kondo, M.~Hotta, Y.~Takagi, Y.~Masukawa,
  T.~Kitahara, Y.~Takema, S.~Koyano, S.~Yamazaki, and A.~Hatamochi,
 {\em J. Invest. Derm.}, {\bf 130}, 2511 (2010).

[6] Y.~Masukawa, H.~Narita, H.~Sato, A.~Naoe, N.~Kondo, Y.~Sugai, T.~Oba, R.~Homma,
  J.~Ishikawa, Y.~Takagi, and T.~Kitahara,
 {\em J. Lipid. Res.}, {\bf 50}, 1708 (2009).

[7] D.~van~der Spoel, E.~Lindahl, B.~Hess, G.~Groenhof, A.~E. Mark, and H.~J.~C.
  Berendsen.
 Gromacs: Fast, flexible and free.
 {\em J. Comp. Chem.}, {\bf 26}, 1701 (2005).

[8] D.~van~der Spoel, E.~Lindahl, B.~Hess, A.~R. van Buuren, E.~Apol, P.~J.
  Meulenhoff, D.P. Tieleman, A.~L. T.~M. Sijbers, K.~A. Feenstra, R.~van
  Drunen, and H.~J.~C. Berendsen,
 {\em Gromacs User Manual version 3.3}.
 www.gromacs.org (2005).

[9] S.W. Chiu, M.~Clark, V.~Balaji, S.~Subramaniam, H.L. Scott, and E.~Jackobsson,
 {\em Biophys. J.}, {\bf 69}, 1230 (1995).

[10] O.~Berger, O.~Edholm, and F.~J\"ahnig,
 {\em Biophys. J.}, {\bf 72}, 2002 (1997).

[11] C.~Das, M.~Noro, and P.~D. Olmsted,
 {\em Biophys. J.}, {\bf 97}, 1941 (2009).

[13] C.~Das, P.~D. Olmsted, and M.~G. Noro,
 {\em SoftMatter}, {\bf 5}, 4549 (2009).

[14] A.~Weerheim and M.~Ponec,
 {\em Arch. Derm. Res.}, {\bf 293}, 191, (2001).

[15] H.~Farwanah, J.~Wohlrab, R.~H.~H. Neubert, and K.~Raith,
 {\em Anal. Bioanal. Chem.}, {\bf 383}, 632 (2005).

[16] J.~A.\ Bouwstra, A.~de~Graaff, G.~S.\ Gooris, J.~Nijsse, J.~W.\ Wiechers, and
  A.~C. van Aelst,
 {\em J. Invest. Derm.}, {\bf 120}, 750 (2003).

[17] E.\ Mombelli, R.\ Morris, W.\ Taylor, and F.\ Fraternali,
 {\em Biophys. J.}, {\bf 84}, 1507 (2003).

[18] J.\ N.\ Israelachvili,
 {\em Intermolecular and surface forces}.
 (Academic Press, London, 1991).

[19] J.\ Shah, J.\ M.\ Atienza, R.\ I.\ Duclos~Jr., A.\ V.\ Rawlings,
  Z.\ Dong, and G.\ G.\ Shipley,
 {\em J. Lipid Res.}, {\bf 36}, 1936 (1995).

[20] J.\ M.\ Seddon,
 {\em Biochim. Biophys. Acta}, {\bf 1031}, 1 (1990).

\end{document}